\documentstyle[aas2pp4]{article}

\input epsf.sty
\def\spose#1{\hbox to 0pt{#1\hss}}
\def\simlt{\mathrel{\spose{\lower 3pt\hbox{$\mathchar"218$}}
        \raise 2.0pt\hbox{$\mathchar"13C$}}}
\def\simgt{\mathrel{\spose{\lower 3pt\hbox{$\mathchar"218$}}
     \raise 2.0pt\hbox{$\mathchar"13E$}}}
\newcommand{\vect}[1]{{\bf #1}}
\newcommand{\hhat}[1]{\hat{\bf #1}}

\begin{document}

\hyphenation{an-iso-tro-pies}
\hyphenation{quad-ru-pole}
\title{A CMB Polarization Primer}

\author{Wayne Hu\footnote{Alfred P. Sloan Fellow}}
\affil{Institute for Advanced Study, Princeton, NJ 08540}

\author{Martin White}
\affil{Enrico Fermi Institute, University of Chicago,
		Chicago, IL 60637}

\vskip 1truecm
\begin{center}
\leavevmode
\epsfxsize=2.5in \epsfbox{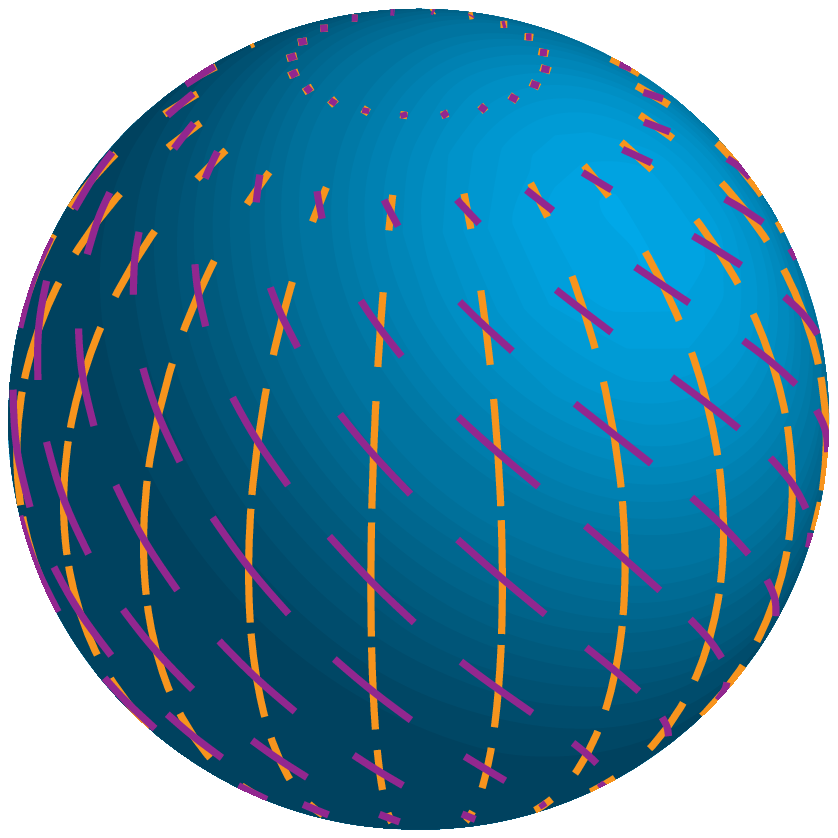}
\end{center}

\begin{abstract}
We present a pedagogical and phenomenological introduction to the study
of cosmic microwave background (CMB) polarization to build intuition
about the prospects and challenges facing its detection.
Thomson scattering of temperature anisotropies on the last
scattering surface generates a linear polarization pattern on the sky that
can be simply read off from their quadrupole moments.   These in turn
correspond directly to the fundamental scalar (compressional), vector
(vortical), and tensor (gravitational wave) modes of cosmological
perturbations.   We explain the origin and phenomenology of the geometric
distinction between these patterns in terms of the so-called electric and
magnetic parity modes, as well as their correlation with the temperature
pattern. By its isolation of the last scattering surface and the
various perturbation modes, the polarization provides unique 
information for the
phenomenological reconstruction of the cosmological model.  
Finally we
comment on the comparison of theory with experimental data
and prospects for the future detection of CMB polarization.
\end{abstract}


\section{Introduction}

Why should we be concerned with the 
polarization of the cosmic microwave
background (CMB) anisotropies?   That the CMB anisotropies are polarized
is a fundamental prediction of the gravitational instability paradigm.
Under this paradigm, small fluctuations in the early universe
grow into the large scale structure we see today. 
If the temperature anisotropies we observe are indeed the result 
of primordial fluctuations, 
their presence at last scattering would polarize the
CMB anisotropies themselves.
The verification of the (partial) polarization of the CMB on small scales would
thus represent a fundamental check on our basic assumptions about the
behavior of fluctuations in the 
universe, in much the same way that the
redshift dependence of the CMB temperature is a test of our assumptions about
the background cosmology.

Furthermore, observations of polarization provide an important tool for
reconstructing the model of the fluctuations from the observed power spectrum
(as distinct from fitting an {\it a priori} 
model prediction to the observations).
The polarization probes the epoch of last scattering {\it directly\/} as
opposed to the temperature fluctuations which may evolve between last
scattering and the present.  This localization in time is a very powerful
constraint for reconstructing the sources of anisotropy.
Moreover, different sources of temperature anisotropies (scalar, vector and
tensor) give different patterns in the polarization: both in its intrinsic
structure and in its correlation with the temperature fluctuations themselves.
Thus by including polarization information, one can distinguish the
ingredients which go to make up the temperature power spectrum and so the
cosmological model.

Finally, the polarization power spectrum provides information complementary
to the temperature power spectrum even for ordinary (scalar or density)
perturbations.  
This can be of use in breaking parameter degeneracies and thus constraining
cosmological parameters more accurately.  The prime example of this is the
degeneracy, within the limitations of cosmic variance, between a change in
the normalization and an epoch of ``late'' reionization.

Yet how polarized are the fluctuations?  
The degree of linear polarization is directly related to the quadrupole
anisotropy in the photons when they last scatter.
While the exact properties of the polarization depend on the mechanism for
producing the anisotropy, several general properties arise.
The polarization peaks at angular scales smaller than the horizon at last
scattering due to causality.
Furthermore, the polarized fraction of the temperature anisotropy is small
since only those photons that last scattered in an optically thin region
could have possessed a quadrupole anisotropy.  
The fraction depends on the duration of last scattering.
For the standard thermal history, it is $10\%$ on a characteristic scale of
tens of arcminutes.
Since temperature anisotropies are at the $10^{-5}$ level, the polarized
signal is at (or below) the $10^{-6}$ level, or several $\mu $K, representing
a significant experimental challenge. 
 
Our goal here is to provide physical intuition for these issues.
For mathematical details, we refer the reader to 
\cite{KamKosSte} (1997), \cite{ZalSel} (1997), \cite{TAMM} (1997)
as well as pioneering work by \cite{BonEfs} (1984) and \cite{Pol} (1985).
The outline of the paper is as follows.  We begin in \S\ref{sec:thomson}
by examining the general properties of polarization formation from Thomson
scattering of scalar, vector and tensor anisotropy sources.
We discuss the properties of the resultant polarization patterns on the
sky and their correlation with temperature patterns in \S\ref{sec:patterns}.
These considerations are applied to the reconstruction problem in
\S\ref{sec:reconstruction}.
The current state of observations and techniques for data analysis are
reviewed in \S \ref{sec:phenomenology}.
We conclude in \S \ref{sec:future} with comments on future prospects for
the measurement of CMB polarization.  

\begin{figure}[ht]
\begin{center}
\leavevmode
\epsfxsize=3.0in \epsfbox{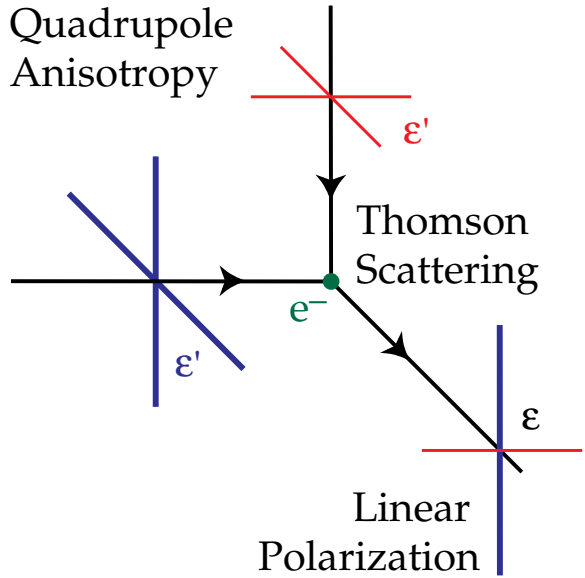}
\end{center}
\caption{Thomson scattering of radiation with a quadrupole anisotropy
generates linear polarization.  Blue colors (thick lines)
represent hot and red colors (thin lines)
cold radiation.  }
\label{fig:thomson}
\end{figure}

\section{Thomson Scattering}
\label{sec:thomson}

\subsection{From Anisotropies to Polarization}
\label{sec:scattering}

The Thomson scattering cross section depends on polarization as (see 
e.g.~\cite{Cha} 1960)
\begin{equation}
  {d\sigma_T\over d\Omega} \propto |\hat{\epsilon}
  \cdot \hat{\epsilon}'|^2\,,
\end{equation}
where $\hat{\epsilon}$ ($\hat{\epsilon}'$) are the incident (scattered)
polarization directions.  Heuristically, the incident light sets up
oscillations of the target electron in the direction of the electric
field vector $\vect{E}$, i.e.~the polarization.
The scattered radiation intensity thus peaks in the direction normal to,
with polarization parallel to, the incident polarization.
More formally, the polarization dependence of the cross section is
dictated by electromagnetic gauge invariance and thus follows from very
basic principles of fundamental physics.

If the incoming radiation field were isotropic, orthogonal polarization
states from incident directions separated by $90^\circ$ would balance so
that the outgoing radiation would remain unpolarized.   Conversely,
if the incident radiation field possesses a {\it quadrupolar\/} variation
in intensity or temperature (which possess intensity peaks at $90^\circ=\pi/2$
separations), the result is a {\it linear\/} polarization of the scattered
radiation (see Fig.~\ref{fig:thomson}).  
A reversal in sign of the temperature fluctuation corresponds to a
$90^\circ$ rotation of the polarization, which reflects the spin-2
nature of polarization.

In terms of a multipole decomposition of the radiation field into spherical
harmonics, $Y_\ell^m (\theta,\phi)$, the five quadrupole moments are
represented by $\ell=2$, $m=0,\pm 1,\pm 2$.
The orthogonality of the spherical harmonics guarantees that no other moment
can generate polarization from Thomson scattering.
In these spherical coordinates, with the north pole at $\theta=0$, we call
a N-S (E-W) polarization component $Q>0$ ($Q<0$) and a NE-SW (NW-SE) component
$U>0$ ($U<0$).  
The polarization amplitude and angle clockwise from north are
\begin{equation}
  P=\sqrt{Q^2+U^2}, \qquad \alpha = {1 \over 2} \tan^{-1}(U/Q) \, .
\end{equation}
Alternatively, the Stokes parameters $Q$ and $U$ represent the diagonal and
off diagonal components of the symmetric, traceless, $2\times 2$ intensity
matrix 
in 
the polarization plane spanned by ($\hhat{e}_\theta$, $\hhat{e}_\phi$),
\begin{equation}
E_i^* E_j -{1\over 2}\delta_{ij} E^2 \propto Q\sigma_3 + U\sigma_1 \, ,
\end{equation}
where $\sigma_i$ are the Pauli matrices and circular polarization
is assumed absent.

If Thomson scattering is rapid, then the randomization of photon directions
that results destroys any quadrupole anisotropy and polarization.
The problem of understanding the polarization pattern of the CMB thus reduces
to understanding the quadrupolar temperature fluctuations at {\it last\/}
scattering.

Temperature perturbations have 3 geometrically distinct sources:
the scalar (compressional), vector (vortical) and tensor (gravitational wave)
perturbations.
Formally, they form the irreducible basis of the symmetric metric tensor.
We shall consider each of these below and show that the 
scalar, vector, and tensor quadrupole anisotropy correspond to
$m=0,\pm 1,\pm 2$ respectively.  
This leads to different patterns of polarization for the
three sources as we shall discuss in \S\ref{sec:patterns}.

\begin{figure}[ht]
\begin{center}
\leavevmode
\epsfxsize=3.25in \epsfbox{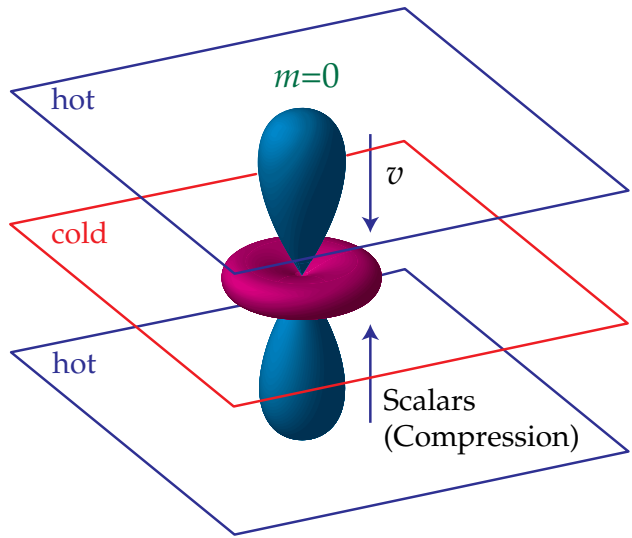}
\end{center}
\caption{The scalar quadrupole moment ($\ell=2,m=0$).  
Flows from hot (blue) regions into cold (red), $\vect{v} \parallel \vect{k}$,
produce the azimuthally symmetric pattern $Y_2^0$ depicted here.}
\label{fig:scalarquad}
\end{figure}

\begin{figure*}[t]
\begin{center}
\leavevmode
\epsfxsize=6in \epsfbox{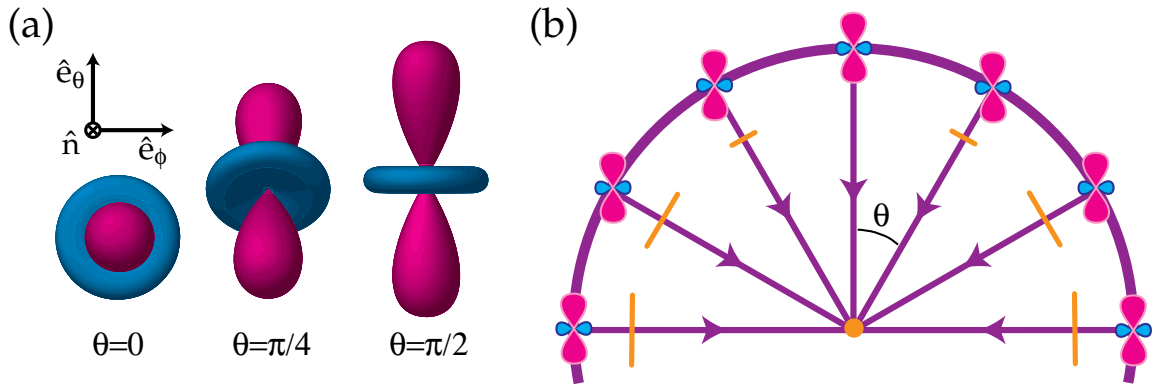}
\end{center}
\caption{The transformation of quadrupole anisotropies into 
linear polarization.
(a) The orientation of the quadrupole moment with respect to
the scattering direction $\hhat{n}$ determines the sense and
magnitude of the polarization.  It is aligned with the 
cold (red, long) lobe in the $\hhat{e}_\theta \otimes \hhat{e}_\phi$ 
tangent plane.  (b) In spherical coordinates where  $\hhat{n} 
\cdot \hhat{k} = \cos\theta$, the polarization points north-south
($Q$) with magnitude varying as $\sin^2\theta$ for scalar 
fluctuations. }
\label{fig:quadtopol}
\end{figure*}

\subsection{Scalar Perturbations}
\label{sec:scalar}

The most commonly considered and familiar types of perturbations
are scalar modes.
These modes represent perturbations in the (energy) density of the
cosmological fluid(s) at last scattering and are the only fluctuations
which can form structure though gravitational instability.

Consider a single large-scale Fourier component of the fluctuation,
i.e.~for the photons, a single plane wave in the temperature perturbation.
Over time, the temperature and gravitational potential gradients cause a
bulk flow, or dipole anisotropy, of the photons.
Both effects can be described by introducing an ``effective'' temperature 
\begin{equation}
  (\Delta T/T)_{\rm eff}= \Delta T/T+\Psi \, ,
\end{equation} 
where $\Psi$ is the gravitational
potential.  Gradients in the effective temperature always
create flows from hot to cold effective temperature.  Formally,
both pressure and gravity act as sources of the momentum density 
of the fluid in a combination that is exactly the effective temperature
for a relativistic fluid.

To avoid confusion, let us explicitly consider the case of adiabatic
fluctuations, 
where initial perturbations to the density imply potential
fluctuations that dominate at large scales.  
Here gravity overwhelms pressure in overdense regions causing matter to
flow towards density peaks initially.
Nonetheless, overdense regions are effectively {\it cold\/} initially
because photons must climb out of the potential wells they create and
hence lose energy in the process.  
Though flows are established from cold to hot temperature regions on large
scales, they still go from hot to cold {\it effective\/} temperature regions.
This property is true more generally of our adiabatic assumption:
we hereafter refer only to effective temperatures to keep the argument
general.

Let us consider the {\it quadrupole\/} component of the temperature pattern
seen by an observer located in a trough of a plane wave.
The azimuthal symmetry in the problem requires that
$\vect{v} \parallel \vect{k}$ and hence the flow is irrotational
$\vect{\nabla} \times \vect{v} =0$.
Because hotter photons from the crests flow into the trough from the
$\pm\hhat{k}$ directions
while cold photons surround the observer in the plane, 
the quadrupole pattern seen in a
trough has an $m=0$, 
\begin{equation} 
Y_2^0 \propto 3 \cos^2\theta - 1\, ,
\end{equation}
structure
with angle $\hhat{n} \cdot \hhat{k} = \cos\theta$
(see Fig.~\ref{fig:scalarquad}).  
The opposite effect occurs at the crests, reversing the sign of the
quadrupole but preserving the $m=0$ nature in its local angular dependence.
The full effect is thus described by a local quadrupole modulated by a plane
wave in space, $- Y_2^0(\hhat{n}) \exp(i \vect{k} \cdot \vect{x})$, 
where the sign denotes the fact that photons flowing into cold regions are
hot.  This infall picture must be modified slightly on scales smaller than
the sound horizon where pressure plays a role (see \S \ref{sec:smallcor}),
however the essential property that the flows are parallel to $\hhat{k}$ and
thus generate an $m=0$ quadrupole remains true.

\begin{figure*}[t]
\begin{center}
\leavevmode
\epsfxsize=6in \epsfbox{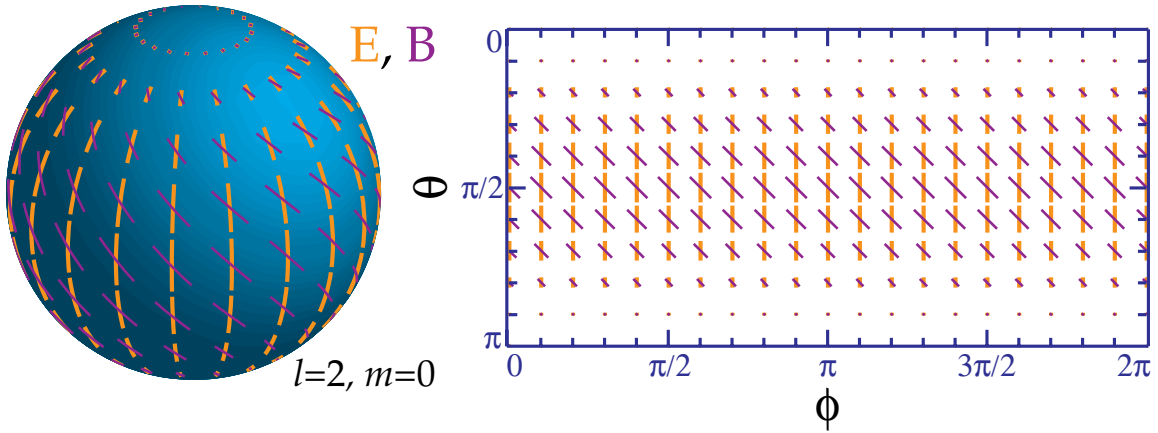}
\end{center}
\caption{Polarization pattern for $\ell=2$, $m=0$, note the azimuthal
symmetry. The scattering of a scalar $m=0$ quadrupole perturbation generates
the electric $E$ (yellow, thick lines) pattern on the sphere.
Its rotation by $45^\circ$ represents the orthogonal magnetic $B$ (purple,
thin lines) pattern.  {\bf Animation} (available at
{\tt http://www.sns.ias.edu/$\sim$whu/polar/scalaran.html}): 
as the line of sight $\hhat{n}$ changes, the lobes of the quadrupole rotate in
and out of the tangent plane.  The polarization follows the orientation of the 
colder (red) lobe in the tangent plane.}
\label{fig:scalarmap}
\end{figure*}

The sense of the quadrupole moment determines the polarization pattern through
Thomson scattering. 
Recall that polarized scattering peaks when the temperature varies in the
direction orthogonal to $\hhat{n}$.  
Consider then the tangent plane $\hhat{e}_\theta \otimes \hhat{e}_\phi$ with
normal $\hhat{n}$.  This may be visualized in an angular ``lobe'' diagram
such as Fig.~\ref{fig:scalarquad} as a plane which passes through the
``origin'' of the quadrupole pattern perpendicular to the line of sight.
The polarization is maximal when the hot and cold lobes of the quadrupole
are in this tangent plane, and is aligned with the component of the colder
lobe which lies in the plane.
As $\theta$ varies from $0$ to $\pi/2$ (pole to equator) the temperature
differences in this plane increase from zero (see Fig.~\ref{fig:quadtopol}a).
The local polarization at the crest of the temperature perturbation is thus
purely in the N-S direction tapering off in amplitude toward the poles
(see Fig.~\ref{fig:quadtopol}b). 
This pattern represents a pure $Q$-field on the sky whose amplitude varies
in angle as an $\ell=2$, $m=0$ tensor or {\it spin-2} spherical harmonic
\begin{equation}
Q = \sin^2\theta, \qquad  U=0.
\end{equation}
In different regions of space, the plane wave modulation of the
quadrupole can change the sign of the polarization, but not its sense.

This pattern (Fig.~\ref{fig:scalarmap}, yellow lines) is of course not the
only logical possibility for an $\ell=2$, $m=0$ polarization pattern.
Its rotation by $45^\circ$ is also a valid configuration (purple lines).
This represents a pure NW-SE (and by sign reversal NE-SW), or $U$-polarization
pattern.  We return in \S\ref{sec:TE} to consider the geometrical distinction
between the two patterns, the {\it electric\/} and {\it magnetic\/} modes.

\begin{figure}[ht]
\begin{center}
\leavevmode
\epsfxsize=3.25in \epsfbox{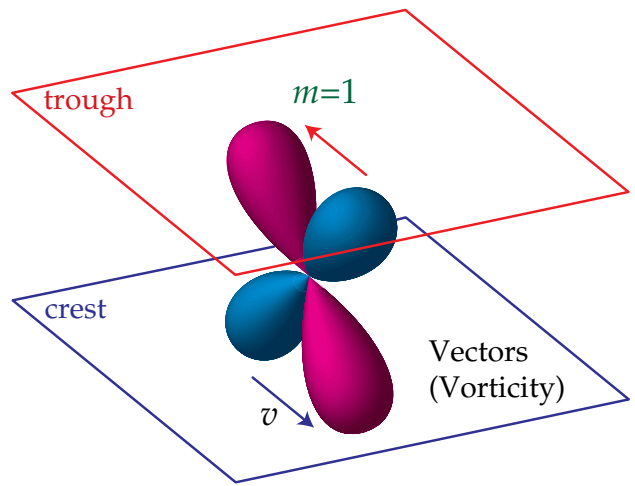}
\end{center}
\caption{The vector quadrupole moment ($\ell=2$, $m=1$).
Since $\vect{v} \perp \vect{k}$, the Doppler effect generates a quadrupole
pattern with lobes $45^\circ$ from $\vect{v}$ and $\vect{k}$ that is spatially
out of phase (interplane peaks) with $\vect{v}$.}
\label{fig:vectorquad}
\end{figure}

\begin{figure*}[t]
\begin{center}
\leavevmode
\epsfxsize=6in \epsfbox{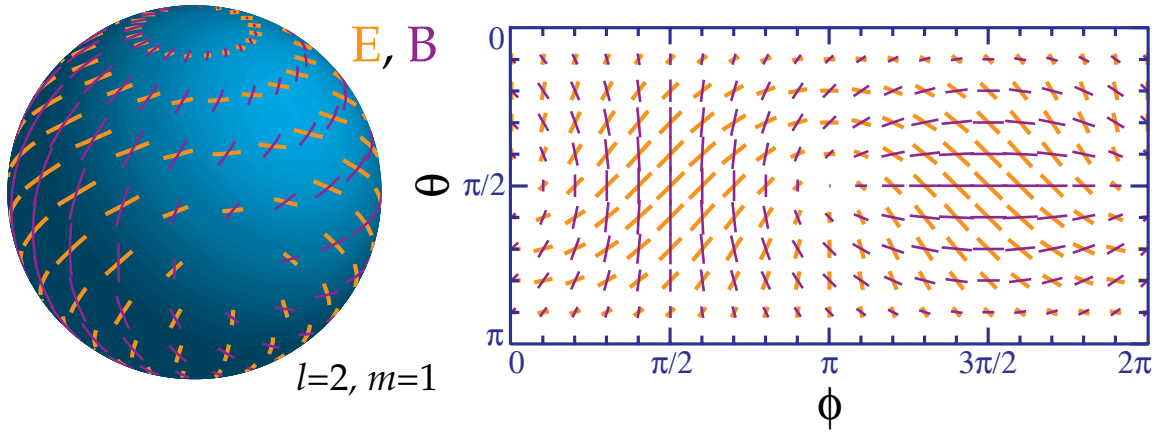}
\end{center}
\caption{Polarization pattern for $\ell=2$, $m=1$.  The scattering of a vector
$(m=1)$ quadrupole perturbation generates the $E$ pattern (yellow, thick lines)
as opposed to the $B$, (purple, thin lines) pattern.  {\bf Animation}
(available at {\tt http://www.sns.ias.edu/$\sim$whu/polar/vectoran.html}):
same as for scalars.}
\label{fig:vectormap}
\end{figure*}

\subsection{Vector Perturbations}
\label{sec:vectors}

Vector perturbations represent vortical motions of the matter, where
the velocity field $\vect{v}$ obeys $\vect{\nabla}\cdot\vect{v}=0$ and $\vect{\nabla}\times\vect{v}\ne0$,
similar to ``eddies'' in water.
There is no associated density perturbation, and the vorticity is
damped by the expansion of the universe as are all motions that are not
enhanced
by gravity.  However, the associated temperature fluctuations, once
generated, do not decay as both $\Delta T$ and $T$ scale similarly with
the expansion.
For a plane wave perturbation, the velocity field $\vect{v}\perp\vect{k}$
with direction reversing in crests and troughs 
(see Fig.~\ref{fig:vectorquad}).  The radiation field at these
extrema possesses a dipole pattern due to the Doppler shift from
the bulk motion.  Quadrupole variations vanish here but peak between
velocity extrema.  
To see this, imagine sitting between crests and troughs.
Looking up toward the trough, one sees the dipole pattern projected
as a hot and cold spot across the zenith; looking down toward the
crest, one sees the projected dipole reversed.  The net effect is
a quadrupole pattern in temperature with $m=\pm 1$
\begin{equation}
Y_2^{\pm 1} \propto \sin\theta \cos\theta e^{\pm i\phi}.
\end{equation}
The lobes are oriented at 45$^\circ$ from $\vect{k}$ and $\vect{v}$ since
the line of sight velocity vanishes along $\vect{k}$ and at 90 degrees to
$\vect{k}$ here.  The latter follows since midway between the crests and
troughs $\vect{v}$ itself is zero.
The full quadrupole distribution is therefore described by
$-i Y_2^{\pm 1}(\hhat{n}) \exp(i \vect{k} \cdot \vect{x})$,
where $i$ represents the spatial phase shift of the quadrupole with
respect to the velocity.

Thomson scattering transforms the quadrupole temperature anisotropy
into a local polarization field as before.  
Again, the pattern may be visualized from the intersection of the
tangent plane to $\hhat{n}$ with the lobe pattern of
Fig.~\ref{fig:vectorquad}.  
At the equator ($\theta=\pi/2$),
the lobes are oriented $45^\circ$ from the line of sight $\hhat{n}$
and rotate into and out of the tangent plane with $\phi$.
The polarization pattern here is a pure $U$-field which varies 
in magnitude as $\sin\phi$.
At the pole $\theta=0$, there are no temperature variations in the
tangent plane so the polarization vanishes.  
Other angles can equally well be visualized by viewing the
quadrupole pattern at different orientations given by $\hhat{n}$.

The full $\ell=2$, $m=1$ pattern, 
\begin{equation}
Q= -\sin\theta \cos\theta e^{i\phi}, \qquad
U= -i \sin\theta e^{i\phi}
\end{equation}
is displayed explicitly in Fig.~\ref{fig:vectormap} (yellow lines, real part).
Note that the pattern is dominated by $U$-contributions especially near
the equator.  The similarities and differences with the scalar pattern
will be discussed more fully in \S\ref{sec:patterns}.

\begin{figure}[ht]
\begin{center}
\leavevmode
\epsfxsize=3.25in \epsfbox{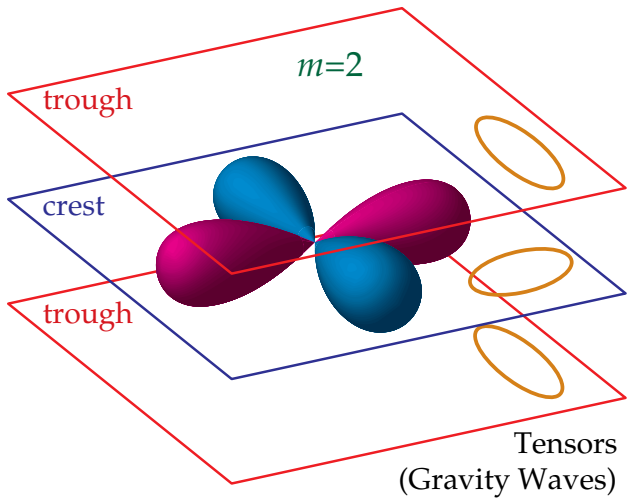}
\end{center}
\caption{The tensor quadrupole moment $(m=2)$. Since gravity waves distort
space in the plane of the perturbation, changing a circle of test particles
into an ellipse, the radiation acquires an $m=2$ quadrupole moment.}
\label{fig:tensorquad}
\end{figure}

\subsection{Tensor Perturbations}
\label{sec:tensors}

Tensor fluctuations are transverse-traceless perturbations to the metric,
which can be viewed as gravitational waves.  A plane gravitational wave
perturbation represents a quadrupolar ``stretching'' of space in the plane
of the perturbation (see Fig.~\ref{fig:tensorquad}).
As the wave passes or its amplitude changes, a circle of test particles in
the plane is distorted into an ellipse whose semi-major axis $\rightarrow$
semi-minor axis as the spatial phase changes from crest $\rightarrow$
trough (see Fig.~\ref{fig:tensorquad}, yellow ellipses).
Heuristically, the accompanying stretching of the wavelength of photons
produces a quadrupolar temperature variation with an $m=\pm 2$ pattern
\begin{equation}
Y_2^{\pm 2} \propto \sin^2\theta e^{\pm 2i\phi}
\end{equation}
in the coordinates defined by $\hhat{k}$.  

\begin{figure*}[t]
\begin{center}
\leavevmode
\epsfxsize=6in \epsfbox{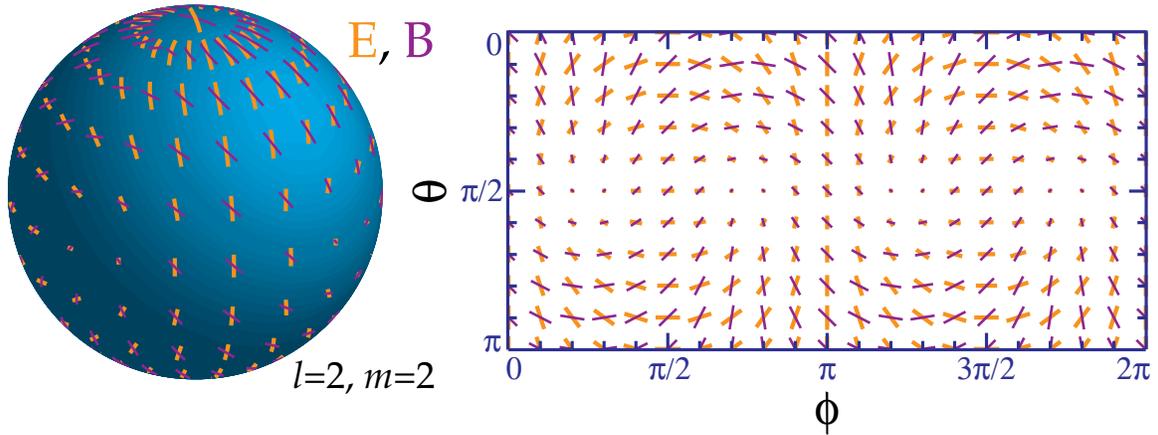}
\end{center}
\caption{Polarization pattern for $\ell=2$ $m=2$. Scattering of a tensor
$m=2$ perturbation generates the $E$ (yellow, thick lines) pattern as
opposed to the $B$ (purple, thin lines) pattern.
{\bf Animation} (available at
{\tt http://www.sns.ias.edu/$\sim$whu/polar/tensoran.html}):
same as for scalars and vectors.}
\label{fig:tensormap}
\end{figure*}

Thomson scattering again produces a polarization pattern from the quadrupole
anisotropy.  At the equator, the quadrupole pattern intersects the tangent
($\hhat{e}_\theta\otimes\hhat{e}_\phi$) plane with hot and cold lobes rotating
in and out of the $\hhat{e}_\phi$ direction with the azimuthal angle $\phi$.
The polarization pattern is therefore purely $Q$ with a $\cos(2\phi)$
dependence.
At the pole, the quadrupole lobes lie completely in the polarization plane
and produces the maximal polarization unlike the scalar and vector cases.
The full pattern,
\begin{equation}
Q = (1+\cos^2\theta)e^{2i\phi}, \qquad 
U = -2i \cos\theta e^{2i\phi},
\end{equation}
is shown in Fig.~\ref{fig:tensormap} (real part).  
Note that $Q$ and $U$ are present in nearly equal amounts for the tensors.

\section{Polarization Patterns}
\label{sec:patterns}

The considerations of \S \ref{sec:thomson} imply that scalars, vectors,
and tensors generate distinct patterns in the polarization of the CMB.  
However, although they separate cleanly into $m=0,\pm 1,\pm 2$ polarization
patterns for a {\it single\/} plane wave perturbation in the coordinate
system referenced to $\vect{k}$, in general there will exist a spectrum of
fluctuations each with a different $\vect{k}$.  
Therefore the polarization pattern  on the sky does not separate into
$m=0,\pm 1,\pm 2$ modes.  In fact, assuming statistical isotropy, one expects
the ensemble averaged power for each multipole $\ell$ to be independent of $m$.
Nonetheless, certain properties of the polarization patterns discussed in the
last section do survive superposition of the perturbations: in particular,
its parity and its correlation with the temperature fluctuations.
We now discuss how one can describe polarization patterns on the sky arising
from a spectrum of $\vect{k}$ modes.

\begin{figure*}[t]
\begin{center}
\leavevmode
\epsfxsize=4.5in \epsfbox{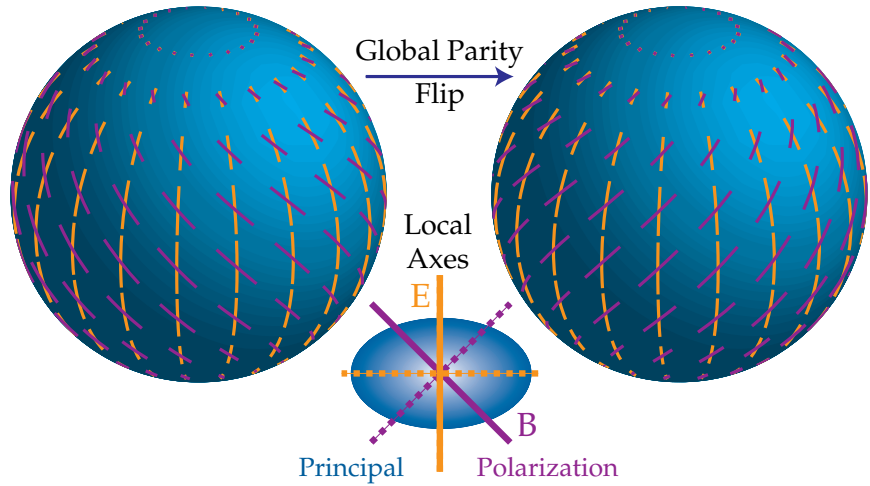}
\end{center}
\caption{The electric ($E$) and magnetic ($B$) modes are distinguished by
their behavior under a parity transformation $\hhat{n} \rightarrow -\hhat{n}$.
$E$-modes have $(-1)^{\ell}$ and $B$-modes have $(-1)^{\ell+1}$ parity;
here $(\ell=2, m=0)$, even and odd respectively.
The local distinction between the two is that the polarization axis is
aligned with the principle axes of the polarization amplitude for $E$ and
crossed with them for $B$.
Dotted lines represent a sign reversal in the polarization.}
\label{fig:parity}
\end{figure*}

\subsection{Electric and Magnetic Modes}
\label{sec:EM}

Any polarization pattern on the sky can be separated into ``electric'' ($E$)
and ``magnetic'' ($B$) components\footnote{These components are called the
``grad'' (G) and ``curl'' (C) components by \cite{KamKosSte} (1997).}.
This decomposition is useful both observationally and theoretically, as we
will discuss below.
There are two equivalent ways of viewing the modes that reflect their global
and local properties respectively.
The nomenclature reflects the global property.  Like multipole radiation,
the harmonics of an $E$-mode have $(-1)^\ell$ parity on the sphere, whereas
those of a $B$-mode have $(-1)^{\ell+1}$ parity.
Under $\hhat{n} \rightarrow -\hhat{n}$, the $E$-mode thus remains unchanged
for even $\ell$, whereas the $B$-mode changes sign as illustrated for the
simplest case $\ell=2,m=0$ in Fig.~\ref{fig:parity}
(recall that a rotation by 90$^\circ$ represents a change in sign).
Note that the $E$ and $B$ multipole patterns are $45^\circ$ rotations of each
other, i.e. $Q \rightarrow U$ and $U \rightarrow -Q$.
Since this parity property is obviously rotationally invariant, it will
survive integration over $\hhat{k}$.

The local view of $E$ and $B$-modes involves the second derivatives
of the polarization amplitude ({\it second\/} derivatives because
polarization is a tensor or spin-2 object).
In much the same way that the distinction between electric and magnetic
fields in electromagnetism involves vanishing of gradients or curls
(i.e.~first derivatives) for the polarization there are conditions on the
second (covariant) derivatives of $Q$ and $U$.
For an $E$-mode, the difference in second (covariant) derivatives of $U$
along $\hhat{e}_\theta$ and $\hhat{e}_\phi$ vanishes as does that for $Q$
along $\hhat{e}_\theta+\hhat{e}_\phi$ and $\hhat{e}_\theta-\hhat{e}_\phi$.
For a $B$-mode, $Q$ and $U$ are interchanged.  Recalling that a $Q$-field
points in the $\hhat{e}_\theta$ or $\hhat{e}_\phi$ direction and a $U$-field
in the crossed direction, we see that the Hessian or curvature matrix of
the polarization amplitude has principle axes in the same sense as the
polarization for $E$ and 45$^\circ$ crossed with it for $B$
(see Fig.~\ref{fig:parity}).
Stated another way, near a maximum of the polarization (where the first
derivative vanishes) the direction of greatest change in the polarization
is parallel/perpendicular and at $45^\circ$ degrees to the polarization in
the two cases.

The distinction is best illustrated with examples.  Take the simplest case
of $\ell=2$, $m=0$ where the $E$-mode is a $Q=\sin^2\theta$ field and the
$B$-mode is a $U=\sin^2\theta$ field (see Fig.~\ref{fig:scalarmap}).
In both cases, the major axis of the curvature lies in the $\hhat{e}_\theta$
direction.  For the $E$-mode, this is in the same sense; for the $B$-mode it
is crossed with the polarization direction.  The same holds true for the
$m=1,2$ modes as can be seen by inspection of Fig.~\ref{fig:vectormap} and
\ref{fig:tensormap}. 

\subsection{Electric and Magnetic Spectra} \label{sec:projection}

Thomson scattering can only produce an $E$-mode locally since the spherical
harmonics that describe the temperature anisotropy have $(-1)^\ell$ 
electric parity.
In Figs.~\ref{fig:scalarmap}, \ref{fig:vectormap}, and \ref{fig:tensormap},
the $\ell=2$, $m=0,1,2$ $E$-mode patterns are shown in yellow.  The $B$-mode
represents these patterns rotated by $45^\circ$ and are shown in purple
and cannot be generated by scattering.
In this way, the scalars, vectors, and tensors are similar in that scattering
produces a {\it local\/} $\ell=2$ $E$-mode only.

\begin{figure}[ht]
\begin{center}
\leavevmode
\epsfxsize=3.25in \epsfbox{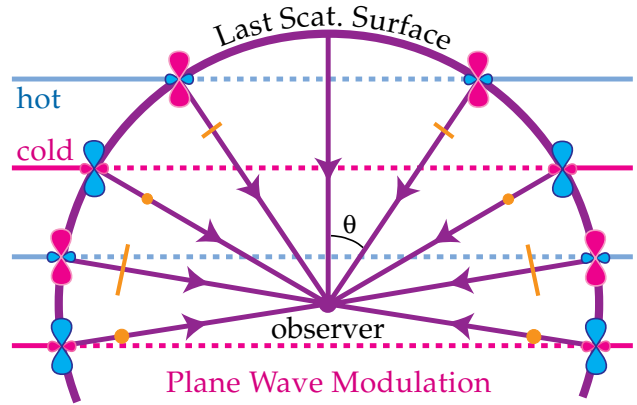}
\end{center}
\caption{Modulation of the local pattern Fig.~\ref{fig:quadtopol}b 
by plane wave fluctuations on the last scattering surface.  
Yellow points represent polarization out of the plane with magnitude
proportional to sign. The plane wave modulation changes the amplitude and
sign of the polarization but does not mix $Q$ and $U$.  Modulation can mix
$E$ and $B$ however if $U$ is also present.}
\label{fig:modulation}
\end{figure}

However, the pattern of polarization on the sky is not simply this local
signature from scattering but is modulated over the last scattering surface
by the plane wave spatial dependence of the perturbation
(compare Figs.~\ref{fig:quadtopol} and \ref{fig:modulation}).
The modulation changes the amplitude, sign, and angular structure of the
polarization but not its nature, e.g. a $Q$-polarization remains $Q$.
Nonetheless, this modulation generates a $B$-mode from the local $E$-mode
pattern.

The reason why this occurs is best seen from the local distinction between
$E$ and $B$-modes.  Recall that $E$-modes have polarization amplitudes that
change parallel or perpendicular to, and $B$-modes in directions $45^\circ$
away from, the polarization direction.  On the other hand, plane wave
modulation always changes the polarization amplitude in the direction
$\hhat{k}$ or N-S on the sphere.  
Whether the resultant pattern possesses $E$ or $B$-contributions depends on
whether the local polarization has $Q$ or $U$-contributions.

\begin{figure*}[t]
\begin{center}
\leavevmode
\epsfxsize=5.75in \epsfbox{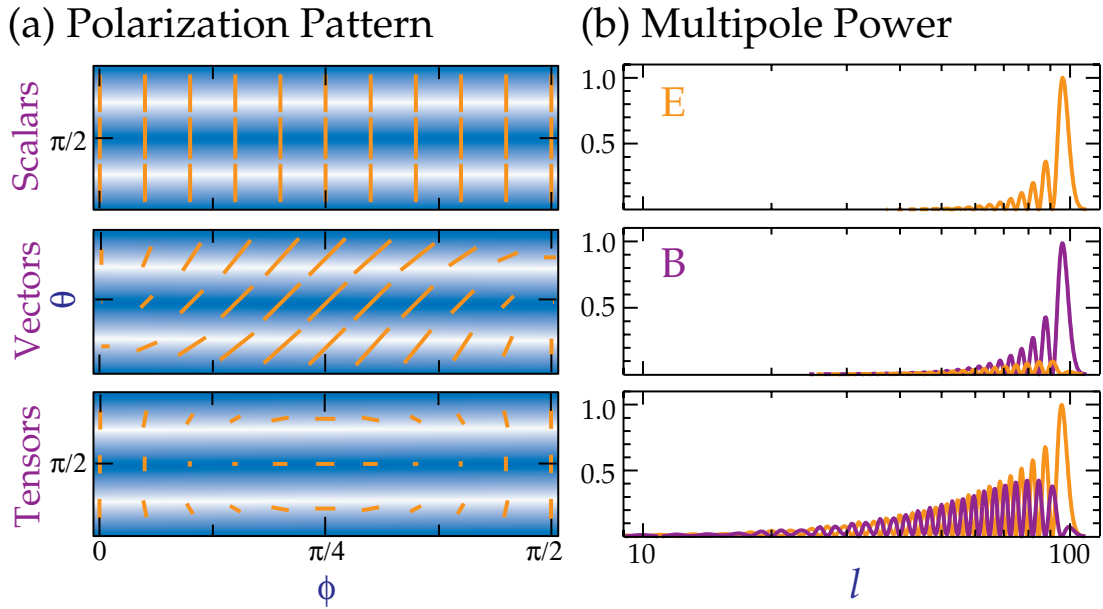}
\end{center}
\caption{The $E$ and $B$ components of a plane wave perturbation.
(a) Modulation of the local $E$-quadrupole pattern (yellow) from scattering
by a plane wave.  Modulation in the direction of (or orthogonal to) the
polarization generates an $E$-mode with higher $\ell$; modulation in the
crossed $(45^\circ)$ direction generates a $B$-mode with higher $\ell$.
Scalars generate only $E$-modes, vectors mainly $B$-modes, and tensors
comparable amounts of both.
(b) Distribution of power in a single plane wave with $kr=100$ in multipole
$\ell$ from the addition of spin and orbital angular momentum.
Features in the power spectrum can be read directly off the pattern in (a).}
\label{fig:plane}
\end{figure*}

For scalars, the modulation is of a pure $Q$-field and thus its $E$-mode
nature is preserved (\cite{KamKosSte} 1997; \cite{ZalSel} 1997).  
For the vectors, the $U$-mode dominates the pattern and the modulation is
crossed with the polarization direction.
Thus vectors generate mainly $B$-modes for short wavelength fluctuations
(\cite{TAMM} 1997).  
For the tensors, the comparable $Q$ and $U$ components of the local pattern
imply a more comparable distribution of $E$ and $B$ modes at short wavelengths
(see Fig.~\ref{fig:plane}a).

These qualitative considerations can be quantified by noting that plane wave
modulation simply represent the addition of angular momentum from the plane
wave ($Y_\ell^0$) with the local spin angular dependence.
The result is that plane wave modulation takes the $\ell=2$ local angular
dependence to higher $\ell$ (smaller angles) and splits the signal into $E$
and $B$ components with ratios which are related to Clebsch-Gordan
coefficients.  At short wavelengths, these ratios are $B/E=0,6,8/13$ in power
for scalars, vectors, and tensors
(see Fig.~\ref{fig:plane}b and \cite{TAMM} 1997).

The distribution of power in multipole $\ell$-space is also important.
Due to projection, a single plane wave contributes to a range of angular
scales $\ell \simlt kr$ where $r$ is the comoving distance to the last
scattering surface.  From Fig.~\ref{fig:modulation}, we see that the smallest
angular, largest $\ell\approx kr$ variations occur on lines of sight
$\hhat{n}\cdot\hhat{k}=0$ or $\theta=\pi/2$ though a small amount of power
projects to $\ell \ll kr$ as $\theta \rightarrow 0$.
The distribution of power in multipole space of Fig.~\ref{fig:plane}b can be
read directly off the local polarization pattern.
In particular, the region near $\theta=\pi/2$ shown in Fig.~\ref{fig:plane}a
determines the behavior of the main contribution to the polarization power
spectrum.  

The full power spectrum is of course obtained by summing these plane wave
contributions with weights dependent on the source of the perturbations and
the dynamics of their evolution up to last scattering.
Sharp features in the $k$-power spectrum will be preserved in the multipole
power spectrum to the extent that the projectors 
in Fig.~\ref{fig:plane}b
approximate delta functions.
For scalar $E$-modes, the sharpness of the projection is enhanced due to strong
$Q$-contributions near $\theta =\pi/2$ ($\ell \sim kr$) that then diminish
as $\theta \rightarrow 0$ $(\ell \ll kr)$.
The same enhancement occurs to a lesser extent for vector 
$B$-modes due to $U$
near $\pi/2$ and tensor $E$-modes due to $Q$ there.  
On the other hand, a supression occurs for vector $E$ and tensor $B$-modes due 
to the
absence of $Q$ and $U$ at $\pi/2$ respectively.  These considerations have
consequences for the sharpness of features in the polarization power spectrum,
and the generation of asymptotic
``tails'' to the polarization spectrum at low-$\ell$ (see \S \ref{sec:adiabaticvisocurvature} and \cite{TAMM} 1997) .

\subsection{Temperature-Polarization Correlation}
\label{sec:TE}

As we have seen in \S \ref{sec:thomson}, the polarization pattern reflects
the local quadrupole anisotropy at last scattering.  Hence the temperature
and polarization anisotropy patterns are correlated in a way that can
distinguish between the scalar, vector and tensor sources.

There are two subtleties involved in establishing the correlation.
First, the quadrupole moment of the temperature anisotropy at last
scattering is not generally the dominant source of anisotropies on the sky,
so the correlation is neither 100\% nor necessarily directly visible as
patterns in the map.  

\begin{figure}[ht]
\begin{center}
\leavevmode
\epsfxsize=3.25in \epsfbox{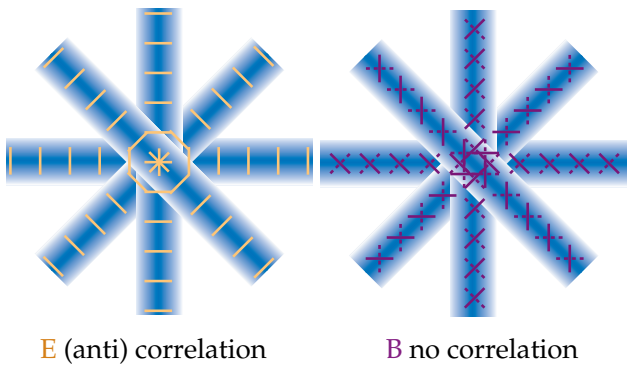}
\end{center}
\caption{Temperature-polarization cross correlation.
$E$-parity polarization perpendicular (parallel) to crests generates
a tangential (radial) polarization field around hot spots. 
$B$-parity polarization does not correlate with temperature
since the $\pm 45^\circ$ rotated contributions from oppositely directed
modes cancel. }
\label{fig:correlation}
\end{figure}

The second subtlety is that the correlation occurs through the $E$-mode
unless the polarization has been Faraday or otherwise rotated between the
last scattering surface and the present.
As we have seen an $E$-mode is modulated in the direction of, or perpendicular
to, its polarization axis.  To be correlated with the temperature, this
modulation must also correspond to the modulation of the temperature
perturbation.  
The two options are that $E$ is parallel or perpendicular to crests in the
temperature perturbation.
As modes of different direction $\hhat{k}$ are superimposed, this translates
into a radial or tangential polarization pattern around hot spots (see
Fig.~\ref{fig:correlation}a).

On the other hand $B$-modes do not correlate with the temperature.
In other words, the rotation of the pattern in Fig.~\ref{fig:correlation}a
by 45$^\circ$ into those of Fig.~\ref{fig:correlation}b (solid and dashed
lines) cannot be generated by Thomson scattering.
The temperature field that generates the polarization has no way to
distinguish between points reflected across the symmetric hot spot and so
has no way to choose between the $\pm 45^\circ$ rotations.  
This does not however imply that $B$ {\it vanishes}.
For example, for a single plane wave fluctuation $B$ can change signs across
a hot spot and hence preserve reflection symmetry
(e.g. Fig.~\ref{fig:vectormap} around the hot spot $\theta=\pi/4$, $\phi=0$).
However superposition of oppositely directed waves as in
Fig.~\ref{fig:correlation}b would destroy the correlation with the hot spot.

The problem of understanding the correlation thus breaks down into 
two steps: (1) determine how the quadrupole moment of the temperature 
at last scattering correlates with the dominant source of 
anisotropies; (2) isolate the $E$-component ($Q$-component 
in $\hhat{k}$ coordinates) and determine whether it represents
polarization parallel or perpendicular to crests and so
radial or tangential to hot spots.

\subsubsection{Large Angle Correlation Pattern}
\label{sec:largecorrelation}

Consider first the large-angle scalar perturbations.  Here 
the dominant source of correlated anisotropies is the 
temperature  perturbation on the last scattering surface itself.
The Doppler contributions can be up to half of the total 
contribution but as we have seen in \S\ref{sec:vectors}
do not correlate with the quadrupole moment.  Contributions
after last scattering, while potentially strong in isocurvature
models for example, also rapidly lose their correlation with 
the quadrupole at last scattering.

As we have seen, the temperature gradient associated with the scalar
fluctuation makes the photon fluid flow from hot regions to cold initially.
Around a point on a crest therefore the intensity peaks in the directions
along the crest and falls off to the neighboring troughs.
This corresponds to a polarization perpendicular to the crest
(see Fig.~\ref{fig:correlation}).
Around a point on a trough the polarization is parallel to the trough.
As we superpose waves with different $\hhat{k}$ we find the pattern is
tangential around hot spots and radial around cold spots (\cite{Crietal} 1995).
It is important to stress that the hot and cold spots refer only to the
temperature component which is correlated with the polarization.
The correlation increases at scales approaching the horizon at last scattering
since the quadrupole anisotropy that generates polarization is caused by flows.

For the vectors, no temperature perturbations exist on the last scattering
surface and again Doppler contributions do not correlate with the quadrupole.
Thus the main correlations with the temperature will come from the quadrupole 
moment itself.
The correlated signal is reduced since the strong $B$-contributions of vectors
play no role.
Hot spots occur in the direction $\theta=\pi/4$, $\phi=0$ where the hot lobe
of the quadrupole is pointed at the observer (see Fig.~\ref{fig:vectorquad}).
Here the $Q$ ($E$) component lies in the N-S direction perpendicular to the
crest (see Fig.~\ref{fig:vectormap}).
Thus the pattern is tangential to hot spot, like scalars
(\cite{TAMM} 1997). 
The signature peaks near the horizon at last scattering for reasons similar
to the scalars.

For the tensors, both the temperature and polarization perturbations arise
from the quadrupole moment, which fixes the sense of the main correlation.
Hot spots and cold spots occur when the quadrupole lobe is pointed at the
observer, $\theta=\pi/2$, $\phi=\pi/2$ and $3\pi/2$.  
The cold lobe and hence the polarization then points in the E-W
direction.  Unlike the scalars and vectors, the pattern will
be mainly radial to hot spots (\cite{Crietal} 1995). 
Again the polarization and hence the cross-correlation peaks near the horizon
at last scattering since gravitational waves are frozen before horizon
crossing (\cite{Pol} 1985). 

\begin{figure}[ht]
\begin{center}
\leavevmode
\epsfxsize=3.25in \epsfbox{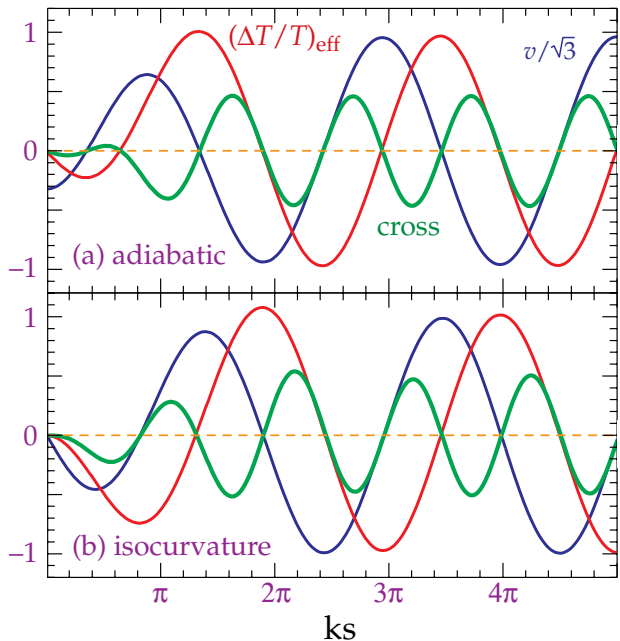}
\end{center}
\caption{Time evolution of acoustic oscillations.  The polarization
is related to the flows $v$ (red) which form quadrupole
anisotropies such that its product (green) with the
effective temperature (blue) reflects the temperature-polarization
cross correlation.
As described in the text the adiabatic (a) and isocurvature (b)
modes differ in the phase of the oscillation in all three quantities.
Temperature and polarization are anticorrelated in both cases
at early times or large scales $ks \ll 1$.}
\label{fig:timeevol}
\end{figure}
\subsubsection{Small Angle Correlation Pattern}
\label{sec:smallcor}

Until now we have implicitly assumed that the
evolution of the perturbations plays a small role as is generally
true
for scales larger than the horizon at last scattering.  
Evolution plays an important role for
small-scale scalar perturbations where there is enough
time for sound to cross the perturbation before last scattering. 
The infall of the photon fluid into troughs compresses the fluid, increasing
its density and temperature.  For adiabatic fluctuations,
this compression reverses the sign of the effective temperature
perturbation when the sound horizon $s$ grows to be $ks \approx \pi/2$
(see Fig.~\ref{fig:timeevol}a).
This reverses the sign of the correlation with the quadrupole moment.
Infall continues until the compression is so great that photon pressure
reverses the flow when $ks \approx \pi$.  Again the correlation reverses sign.
This pattern of correlations and anticorrelations continues
at twice the frequency of the acoustic oscillations themselves 
(see Fig.~\ref{fig:timeevol}a). 
Of course the polarization is only generated at last scattering
so the correlations and anticorrelations are a function of 
scale with sign changes at multiples of $\pi/2 s_*$, where
$s_*$ is the sound horizon at last scattering.  As discussed
in \S\ref{sec:projection}, these fluctuations project onto 
anisotropies as $\ell \sim k r$.

Any scalar fluctuation will obey a similar pattern that reflects
the acoustic motions of the photon fluid.  In particular, at the
largest scales the $ks_* \ll \pi/2 $,  the polarization must 
be anticorrelated with the temperature because the fluid will
always flow with the temperature gradient initially from hot to cold. 
However, where the sign reversals occur depend on the 
acoustic dynamics and so is a useful probe of the nature of
the scalar perturbations, e.g. whether they are adiabatic or
isocurvature (\cite{Signatures} 1996).  In typical isocurvature models, the lack of
initial temperature perturbations delays the acoustic
oscillation by $\pi/2$ in phase
so that correlations reverse at $ks_* = \pi$ (see Fig.~\ref{fig:timeevol}b).   

For the vector and tensor modes, strong evolution can introduce
a small correlation with temperature fluctuations generated {\it after}
last scattering.  The effect is generally weak and model dependent
and so we shall not consider it further here.

\begin{figure*}[tp]
\begin{center}
\leavevmode
\epsfxsize=5.75in \epsfbox{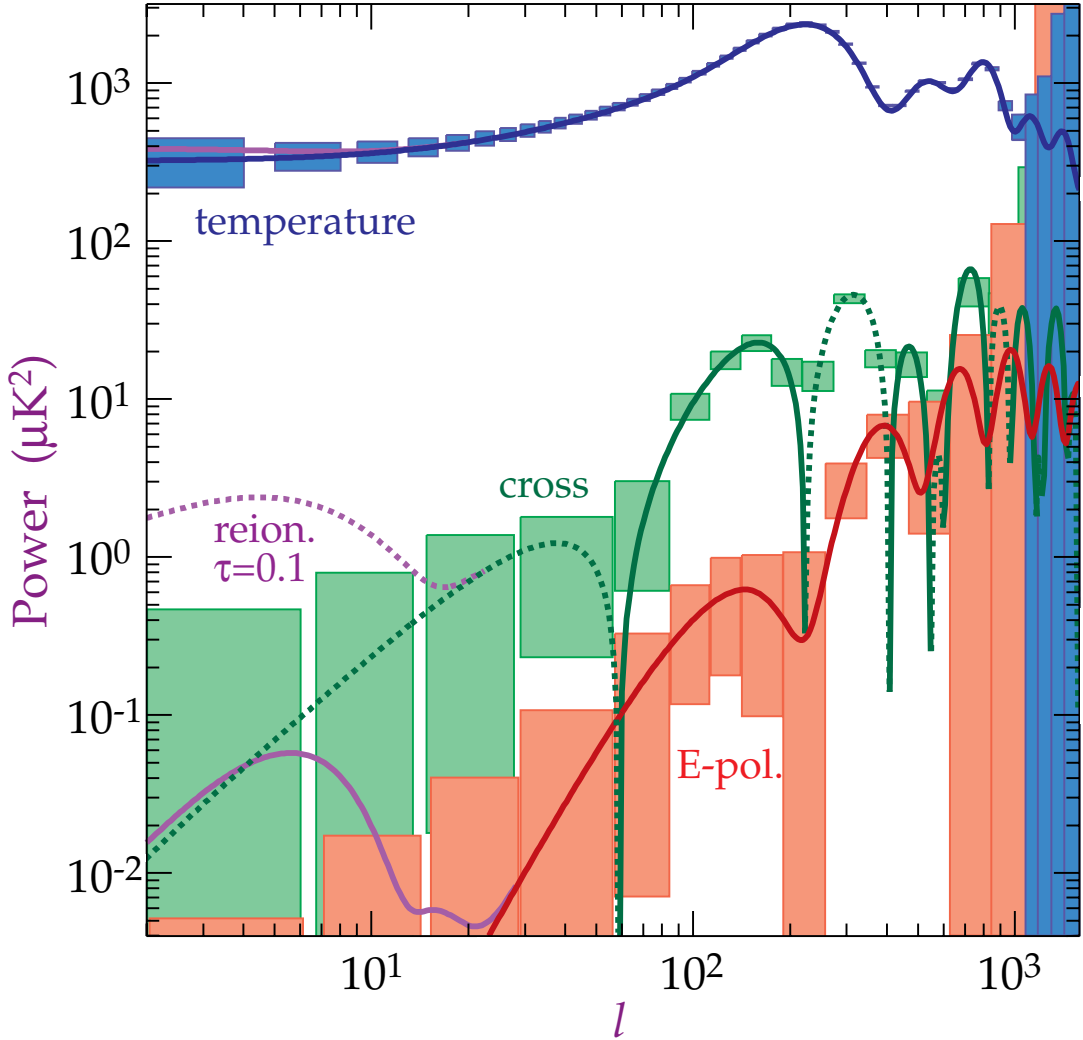}
\end{center}
\caption{Temperature, polarization, and temperature-polarization
cross correlation predictions and sensitivity of MAP for
a fiducial model $\Omega_0=1$, $\Omega_B=0.1$, $h=0.5$ cold
dark matter.  The raw MAP satellite sensitivity (1 $\sigma$ errors
on the recovered power spectrum binned in $\ell$) is approximated
by noise weights of $w_T^{-1} = (0.11\mu K)^2$ for
the temperature and $w_P^{-1} = (0.15 \mu K)^2$ for
the polarization and a FWHM beam of $0.25^\circ$.  Note that errors 
between the spectra are correlated.
While the reionized model (purple $\tau=0.1$)
is impossible to distinguish from the fiducial model from
temperature anisotropies alone, its effect on polarization
is clearly visible at low $\ell$.  Dashed lines for the 
temperature-polarization correlation represent anticorrelation.
{\bf Animation} (available at
{\tt http://www.sns.ias.edu/$\sim$whu/polar/tauan.html}): Variations in
the spectra as $\tau$ is stepped from $0-1$.}
\label{fig:sCDM}
\end{figure*}

\section{Model Reconstruction}
\label{sec:reconstruction}

While it is clear how to compare theoretical predictions of 
a given model with observations, the reconstruction of
a phenomenological model from the data is a more subtle issue.
The basic problem is that in the CMB, we see the whole history
of the evolution in redshift projected onto the two dimensional
sky.  The reconstruction of the evolutionary history of the universe
might thus seem an ill-posed problem.

Fortunately, one needs only to assume the very basic
properties of the cosmological model and the gravitational
instability picture before useful information may be extracted.
The simplest example is the combination of
the amplitude of the temperature fluctuations, which reflect
the conditions at horizon crossing, and large
scale structure today. Another example is the acoustic peaks in the
temperature which form a snapshot of conditions at last
scattering on scales below the horizon at that time.
In most models, the acoustic signature provides a wealth of
information on cosmological parameters and structure formation
(\cite{Signatures} 1996).  Un\-for\-tunate\-ly, it does not directly
tell us the behavior on the largest scales where important 
causal distinctions between models lie.  Furthermore it
may be absent in models with complex evolution on small scales
such as cosmological defect models.  

Here we shall consider how polarization information aids
the reconstruction process by isolating the last scattering
surface on large scales, and separating scalar, vector and tensor
components.  If and when these properties are determined, it will 
become
possible to establish observationally the basic properties 
of the cosmological model such as the nature of the initial fluctuations,
the mechanism of their generation, and the thermal history of the universe.

\subsection{Last Scattering}

The main reason why polarization is so useful to the reconstruction process
is that (cosmologically) it can only be generated by Thomson scattering.
The polarization spectrum of the CMB is thus a direct snapshot of conditions
on the last scattering surface.  Contrast this with temperature fluctuations
which can be generated by changes in the metric fluctuations
between last scattering and the
present, such as those
created by gravitational potential evolution. 
This is the reason why the mere detection of large-angle anisotropies by
{\sl COBE\/} did not rule out wide classes of models such as cosmological
defects.  To use the temperature anisotropies for the reconstruction
problem, one must isolate features in the spectrum which
can be associated with last scattering, or more generally, the
universe at a known redshift.

Furthermore, the polarization spectrum has potential advantages even 
for extracting information from 
features that are also present in the temperature
spectrum, e.g.~the acoustic peaks.  The polarization spectrum is
generated by local quadrupole anisotropies alone whereas the temperature
spectrum has comparable contributions from the local monopole and dipole
as well as possible contributions between last scattering and the present.
This property enhances the prominence of features (see Fig.~\ref{fig:sCDM})
as does the fact that the scalar polarization has a relatively sharp
projection due to the geometry (see \S \ref{sec:projection}).
Unfortunately, these considerations are mitigated by the fact that the
polarization amplitude is so much weaker than the temperature.
With present day detectors, one needs to measure it in broad $\ell$-bands
to increase the signal--to--noise (see Fig.~\ref{fig:sCDM} and
Table~\ref{tab:bandpower}).

\subsection{Reionization}

Since polarization directly probes the last scattering epoch the first thing
we learn is when that occurred, i.e.~what fraction of photons last scattered
at $z\approx1000$ when the universe recombined, and what fraction rescattered
when the intergalactic medium reionized at $z_{\rm ri} \simgt 5$.

Since rescattering erases fluctuations below the horizon scale
and regenerates them only weakly (\cite{Efs} 1988), 
we already know from the
reported excess (over {\sl COBE}) of sub-degree scale anisotropy that the
optical depth during the reionized epoch was $\tau \simlt 1$ and hence 
\begin{equation}
z_{\rm ri} \simlt 100 \left({\Omega_0 h^2 \over 0.25}\right)^{1/3}
\left({\Omega_B h^2 \over 0.0125}\right)^{-2/3} .
\end{equation}  
It is thus likely that our universe has the interesting property
that {\it both\/} the recombination and reionization epoch are
observable in the temperature and polarization spectrum.

Unfortunately for the temperature spectrum, at these low optical depths
the main effect of reionization is an erasure of the primary anisotropies
(from recombination) as $e^{-\tau}$.  This occurs below the horizon at last
scattering, since only on these scales has there been sufficient time to
convert the originally isotropic temperature fluctuations into anisotropies.
The uniform reduction of power at small scales has the same effect as a
change in the overall normalization. 
For $z_{\rm ri}\sim 5$--20 the difference in the power spectrum is confined
to large angles ($\ell<30$).
Here the observations are limited by ``cosmic variance'': the fact that we
only have one sample of the sky and hence only $2\ell+1$ samples of any given
multipole.
Cosmic variance is the dominant source of uncertainty on the low-$\ell$
temperature spectrum in Fig.~\ref{fig:sCDM}.  

The same is not true for the polarization.
As we have seen the polarization spectrum is very sensitive to the epoch of
last scattering.  More specifically, the location of its peak depends on the
horizon size at last scattering and its height depends on the duration of
last scattering (\cite{Efs} 1988).   
This signature is not cosmic variance limited until quite late reionization,
though the combination of low optical depth and partial polarization will
make it difficult to measure in practice (see Fig.~\ref{fig:sCDM} and
Table~\ref{tab:bandpower}).
Optical depths of a few percent 
are potentially observable from the MAP satellite
(\cite{ZalSpeSel}~1997) and of order unity
from the POLAR experiment (\cite{Keaetal} 1997).  

On the other hand, these considerations imply that  the interesting
polarization signatures from recombination will not be completely
obscured by reionization.  We turn to these now.
 
\begin{figure*}[tp]
\begin{center}
\leavevmode
\epsfxsize=5.75in \epsfbox{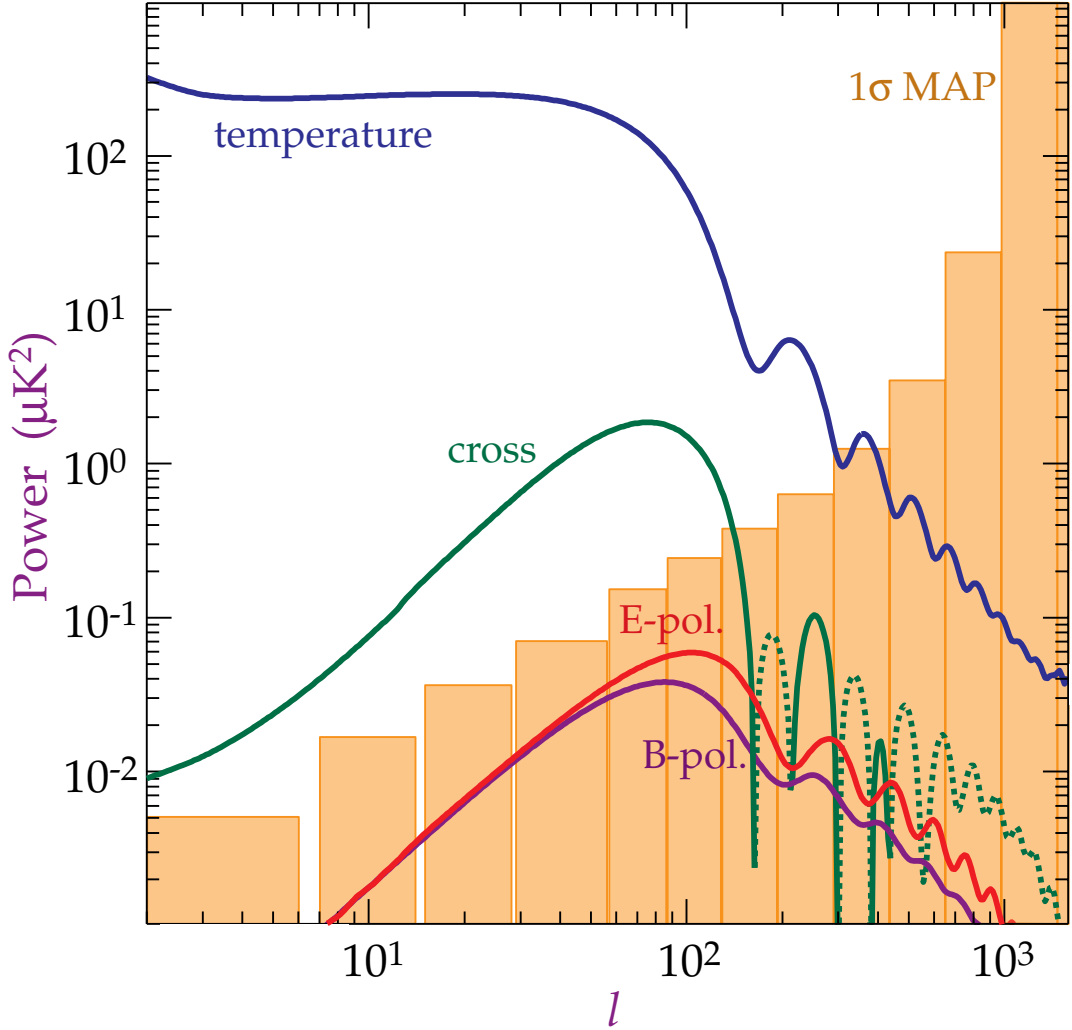}
\end{center}
\caption{Tensor power spectra.  The tensor temperature, 
temperature-cross-polarization, $E$-mode, and $B$-mode
polarization for a scale invariant intial spectrum of
tensors with $\Omega_0=1.0$, $\Omega_B=0.1$ and $h=0.5$.
The normalization has been artificially set (high) to have the
same quadrupole as the scalar spectrum of Fig.~\ref{fig:sCDM}.
Also shown is the MAP $1\sigma$ upper limits on the $B$-mode
due to noise, i.e.~assuming {\it no} signal, showing that
higher sensitivity will likely be necessary to obtain a significant detection 
of $B$-polarization
from tensors.}
\label{fig:tensor}
\end{figure*}

\subsection{Scalars, Vectors, \& Tensors}

There are three types of fluctuations: scalars, vectors and tensors, and
four observables: the temperature, $E$-mode, $B$-mode, and
temperature cross polarization power spectra.
The CMB thus provides sufficient information to separate these contributions,
which in turn can tell us about the generation mechanism for fluctuations in
the early universe (see \S \ref{sec:inflationvdefects}).

Ignoring for the moment the question of foregrounds, to which we turn in
\S\ref{sec:foregrounds}, if the $E$-mode polarization greatly exceeds the
$B$-mode then scalar fluctuations dominate the anisotropy.  
Conversely if the $B$-mode is greater than the $E$-mode, then vectors
dominate.  If tensors dominate, then the $E$ and $B$ are comparable
(see Fig.~\ref{fig:tensor}).
These statements are independent of the dynamics and underlying spectrum
of the perturbations themselves.  

The causal constraint on the generation of a quadrupole moment
(and hence the polarization) introduces further distinctions.
It tells us that the polarization peaks around the scale the horizon subtends
at last scattering.  This is about a degree in a flat universe and scales with
the angular diameter distance to last scattering.
Geometric projection tells us that the low-$\ell$ tails of the polarization
can fall no faster than $\ell^6$, $\ell^4$ and $\ell^2$ for scalars, vectors
and tensors (see \S \ref{sec:projection}).
The cross spectrum falls no more rapidly than $\ell^4$ for each.
We shall see below that these are the predicted slopes of an isocurvature
model.  

Furthermore, causality sets the scale that separates the large and small angle
temperature-polarization correlation pattern.  Well above this scale, scalar
and vector fluctuations should show anticorrelation (tangential around hot
spots) whereas tensor perturbations should show correlations (radial around
hot spots).  

Of course one must use measurements at small enough angular scales that
reionization is not a source of confusion and one must understand the
contamination from foregrounds extremely well.  The latter
is especially true at large angles where the polarization
amplitude decreases rapidly.

\subsection{Adiabatic vs. Isocurvature Perturbations}
\label{sec:adiabaticvisocurvature}

The scalar component is interesting to isolate since it alone is
responsible for large scale structure formation.  
There remain however two possibilities.
Density fluctuations could be present initially.  This represents the
{\it adiabatic\/} mode.  Alternately, they can be generated from stresses
in the matter which causally push matter around.  This represents the
{\it isocurvature\/} mode.  

The presence or absence of density perturbations above the horizon at
last scattering is crucial for the features in both the temperature
and polarization power spectrum.
As we have seen in \S\ref{sec:largecorrelation}, it has 
as strong effect on the phase
of the acoustic oscillation.  In a typical isocurvature model, the phase
is delayed by $\pi/2$ moving structure in the temperature spectrum to
smaller angles (see Fig.~\ref{fig:timeevol}).
Consistency checks exist in the $E$-polarization and cross spectrum which
should be out of phase with the temperature spectrum and oscillating at
twice the frequency of the temperature respectively.  
However, if the stresses are set up sufficiently carefully, this acoustic
phase test can be evaded by an isocurvature model (\cite{Tur} 1996).  
Perhaps more importantly, acoustic features can be  washed
out in isocurvature models with complicated small scale dynamics
to force the acoustic oscillation, as in many defect models
(\cite{Albetal} 1996).

Even in these cases, 
the polarization carries a robust signature of isocurvature
fluctuations.
The polarization isolates the last scattering surface and eliminates any
source of confusion from the epoch between last scattering and the present.
In particular, the delayed generation of density perturbations in these
models implies a steep decline in the polarization above the angle subtended
by the horizon at last scattering.
The polarization power thus hits the asymptotic limits given in the previous
section in contrast to the adiabatic power spectra shown in
Fig.~\ref{fig:sCDM} which have more power at large angles.
To be more specific, if the $E$-power spectrum falls off as $\ell^6$ and or
the cross spectrum as $\ell^4$, then the initial fluctuations are
isocurvature in nature (or an adiabatic model with a large spectral index,
which is highly constrained by {\sl COBE}).

Of course measuring a steeply falling spectrum is difficult in 
practice.  Perhaps more easily measured is the first feature 
in the adiabatic $E$-polarization spectrum at twice the angular
scale of the first temperature peak and the first sign crossing
of the correlation at even larger scales.  
Since isocurvature models must be pushed
to the causal limit to generate these features, their observation
would provide good evidence for the adiabatic nature of
the initial fluctuations (\cite{HuSpeWhi} 1997).

\subsection{Inflation vs. Defects}
\label{sec:inflationvdefects}

One would like to know not only the nature of the fluctuations, but also the
means by which they are generated.  We assume of course that they are not
merely placed by {\it fiat\/} in the initial conditions.
Let us first divide the possibilities into broad classes.
In fact, the distinction between isocurvature and adiabatic fluctuations is
operationally the same as the distinction between conventional causal sources
(e.g.~defects) and those generated by a period of superluminal expansion in
the early universe (i.e.~inflation).
It can be shown that inflation is the only causal mechanism for generating
superhorizon size density (curvature) fluctuations (\cite{Lid} 1995).
Since the slope of the power spectrum in $E$ can be traced directly to the 
presence of ``superhorizon size'' temperature, and hence curvature,
fluctuations at last scattering, it represents a ``test'' of
inflation (\cite{TAMM} 1997, \cite{SpeZal} 1997).  
The acoustic phase test, either in the temperature or polarization,
represents a marginally less robust test that should be easily observable if
the former fails to be.

These tests, while interesting, do not tell us anything about the detailed
physics that generates the fluctuations.  Once a distinction is made between
the two possibilities one would like to learn about the mechanism for
generating the fluctuations in more detail.  For example in the inflationary
case there is a well known test of 
single-field slow-roll inflation which can
be improved by using polarization information.
In principle, inflation generates both scalar and tensor anisotropies.
If we assume that the two spectra come from a single underlying inflationary
potential their amplitudes and slopes are not independent.
This leads to an algebraic consistency relation between the ratio of the
tensor and scalar perturbation spectra and the tensor spectral index.
However information on the tensor contribution to the spectrum is limited by
cosmic variance and is easily confused with other effects such as those of a
cosmological constant or tilt of the initial spectrum.
By using polarization information much smaller ratios of tensor to scalar
perturbations may be probed, with more accuracy, and the test refined
(see \cite{ZalSpeSel} 1997 for details).  
In principle, this extra information may also allow one to reconstruct the low
order derivatives of the inflaton potential.

Similar considerations apply to causal generation of fluctuations 
without inflation.
Two possibilities are a model with ``passive evolution'' where initial stress
fluctuations move matter around and ``active evolution'' where exotic but
causal physics continually generates stress-energy perturbations inside the
horizon.  All cosmological defect models bear aspects of the
latter class. 
The hallmark of an active generation mechanism is the presence of vector
modes.  Vector modes decay with the expansion so in a passive model they would
no longer be present by recombination.
Detection of (cosmological) vector modes would be strong evidence for
defect models.  
Polarization is useful since it provides a unique signature of vector modes
in the dominance of the $B$-mode polarization (\cite{TAMM} 1997).
Its level for several specific defect models is given in  
\cite{SelPenTur} (1997).

\section{Phenomenology}
\label{sec:phenomenology}

\subsection{Observations}
 
While the theoretical case for observing polarization is strong, it is a
difficult experimental task to observe signals of the low level of several
$\mu $K and below.  Nonetheless,  polarization experiments
have one potential advantage over temperature anisotropy experiments.
They can reduce atmospheric emission effects by differencing the polarization
states on the same patch of sky instead of physically chopping between
different angles on the sky since atmospheric emission is thought
to be nearly unpolarized (see \S \ref{sec:foregrounds}).  
However, to be successful an experiment must
overcome a number of systematic effects, many of which are 
discussed in \cite{Keaetal} (1997).  It must at least
balance the sensitivity of the instrument to the 
orthogonal polarization
channels (including the far side lobes) to nearly 
8 orders of magnitude.
Multiple levels of switching and a very careful design are minimum
requirements.

To date the experimental upper limits on polarization of the CMB have been
at least an order of magnitude larger than the theoretical expectations.
The original polarization limits go back to Penzias \& Wilson (1965) who
set a limit of 10\% on the polarization of the CMB.
There have been several subsequent upper limits which have now reached the
level of $\sim20\mu$K (see Tab.~\ref{tab:data}), about a factor of 5--10 above the
predicted levels for popular models (see Tab.~\ref{tab:bandpower}).

\begin{table*}[t]
\begin{center}
\begin{tabular}{lcccc} \hline
Reference & Frequency (GHz) & Ang. Scale & $\Delta T (\mu$K) & (CL) \\ \hline
\cite{PenWil} (1965) & 4.0	& --	& $10^5$	& 95\% \\
\cite{Nan} (1979)		& 9.3	& $15^\circ$ beam & 
	$2\times10^{3}$ & 90\% \\
\cite{Cadetal} (1978)		&100-600&
  $0.5^\circ \simlt \theta \simlt 40^\circ$& 
$2\times10^{3}(40^\circ/\theta)^{1/2}$ & 65\%\\
\cite{LubSmo} (1981)	& 33	& $\ell\la3$ & $200$ & 95\% \\
\cite{Paretal} (1988) &  5    & $1'\simlt\theta\simlt 3'$
	& $100$ & 95\% \\
\cite{Woletal} (1993)   &26-36  & $50 \simlt \ell \simlt 100$ & $25$ & 95\% \\
\cite{Netetal} (1995)      &26-46  & $50 \simlt \ell \simlt 100$ & $16$ & 95\% \\ \hline
\end{tabular}
\end{center}
\caption{Experimental upper limits on the polarization of the CMB. 
The values are converted from the quoted $\Delta T/T$ by multiplying
by $T = 2.728$K, thus these numbers only approximate the bandpowers
used in Tab.~\ref{tab:bandpower}.}
\label{tab:data}
\end{table*}

\begin{table*}
\begin{center}
\begin{tabular}{lccc|lll|lll|lll }\hline
\multicolumn{4}{c|}{Cosmological Parameters} &
\multicolumn{3}{c|}{$\ell$: 2--20 ($\mu$K)} & 
\multicolumn{3}{c|}{$\ell$: 100-200 ($\mu$K)} & 
\multicolumn{3}{c}{$\ell$: 700--1000 ($\mu$K)} \\ 
\multicolumn{1}{c}{$\Omega_0$} & 
\multicolumn{1}{c}{$\Omega_{\rm B}$} & 
\multicolumn{1}{c}{$z_{\rm ri}$} & 
\multicolumn{1}{c|}{$T/S$} &
\multicolumn{1}{c}{$E$} &
\multicolumn{1}{c}{$TE$} &
\multicolumn{1}{c|}{$B$} &
\multicolumn{1}{c}{$E$} &
\multicolumn{1}{c}{$TE$} &
\multicolumn{1}{c|}{$B$} &
\multicolumn{1}{c}{$E$} &
\multicolumn{1}{c}{$TE$} &
\multicolumn{1}{c}{$B$} \\
\hline
1.0 & 0.05 & 0 & 0 &  0.02 & ($-$)0.42 & 0 & 
             0.68  & (+)3.6  & 0 &
	     3.7  & (+)4.8  & 0 \\
1.0 & 0.10 & 0 & 0 & 0.017 & ($-$)0.40 & 0 & 
	    0.71  & (+)4.2  & 0 &
	    3.4   & (+)3.3  & 0 \\
1.0 & 0.05 & 5 & 0 & 0.03 & ($-$)0.59 & 0 &
	         0.68 & (+)3.6  & 0 &
	         3.7  & (+)4.8  & 0 \\
1.0 & 0.05 &10 & 0 & 0.06 & ($-$)0.78 & 0 &  
	         0.68 & (+)3.6  & 0 & 
	         3.6  & (+)4.8  & 0 \\ 
1.0 & 0.05 &20 & 0 & 0.14 & ($-$)1.1 & 0 &  
	         0.66 & (+)3.5  & 0 &
	         3.5  & (+)4.7  & 0 \\
1.0 & 0.05 & 0 & 0.1 & 0.02 & ($-$)0.40 & 0.01 &
		       0.65 & (+)3.4 & 0.05 &
		       3.5  & (+)4.6 & 0.01 \\ 
$0.4_K$ & 0.05 & 0.1 & 0  & 0.01 & ($-$)0.18 & 0 & 
		       0.44 & (+)2.3  & 0 & 
		       2.8  & (+)5.5  & 0  \\
$0.4_\Lambda$ & 0.05 & 0 & 0 & 0.02 & ($-$)0.37 & 0 & 
		       0.65 & (+)4.0  & 0 & 
		       4.1  & (+)6.4  & 0  \\
\hline
\end{tabular}
\end{center}
\caption{Bandpowers for a selection of
cosmological models all with $h=0.5$ and scale invariant spectra save
for the $T/S=0.1$ model where $T/S = 7(1-n_s)$.  The low-$\Omega_0$
models are indicated as $0.4_K$ for the open model and $0.4_\Lambda$ for
the model with $\Omega_\Lambda=1-\Omega_0=0.6$.
The bandpower is defined in units of the COBE quadrupole over the 
range of $\ell$ indicated. The sign of the cross correlation is
indicated in parentheses and note that the band averaged signal may
be suppressed from cancellation of correlated and anticorrelated 
angular regimes.}
\label{tab:bandpower}
\end{table*}

\subsection{Foregrounds}\label{sec:foregrounds}

Given that the amplitude of the polarization is so small the question of
foregrounds is even more important than for the temperature anisotropy.
Unfortunately, the level and structure of the various foreground polarization
in the CMB frequency bands is currently not well known.
We review some of the observations in the adjacent radio and IR bands
(a more complete discussion can be found in \cite{Keaetal} 1997).
Atmospheric emission is believed to be negligibly polarized
(\cite{Keaetal} 1997), leaving the main astrophysical foregrounds:
free-free, synchrotron, dust, and point source emissions.
Of these the most important foreground is synchrotron emission.

Free-free emission (bremsstrahlung) is intrinsically unpolarized
(\cite{RybLig} 1979) but can be partially polarized by Thomson scattering
within the HII region.  This small effect is not expected to polarize the
emission by more than 10\% (\cite{Keaetal} 1997).
The emission is larger at low frequencies but is not expected to dominate
the polarization at any frequency. 

The polarization of dust is not well known.
In principle, emission from dust particles could be highly polarized,
however \cite{HilDra} (1995) find that in their observations
the majority of dust is polarized
at the $\approx 2\%$ level at $100\mu$m with a small fraction of regions
approaching $10\%$ polarization.
Moreover \cite{Keaetal} (1997) show that even at 100\% polarization,
extrapolation of the IRAS 100$\mu$m map with the {\sl COBE\/} FIRAS index
shows that dust emission is negligible below $80$GHz.
At higher frequencies it will become the dominant foreground.

Radio point sources are polarized due to synchrotron emission at 
$<20\%$ level.  For large angle experiments, the random contribution
from point sources will contribute negligibly, but may be of more
concern for the upcoming satellite missions.

Galactic synchrotron emission is the major concern.  It is potentially
highly polarized with the fraction dependent on the spectral index and
depolarization from Faraday rotation and non-uniform magnetic fields.
The level of polarization is expected to lie between 10\%-75\% of a total
intensity which itself is approximately $50\mu $K at $30$GHz.
This estimate follows from extrapolating the \cite{BroSpo} (1976)
measurements at 1411 MHz with an index of $T\propto\nu^{-3}$.

Due to their different spectral indices, the minimum in the foreground
polarization, like the temperature, lies near 100GHz.  
For full sky measurements, since synchrotron emission is more highly
polarized than dust, the optimum frequency at which to measure intrinsic
(CMB) polarization is slightly higher than for the anisotropy.  Over
small regions of the sky where one or the other of the foregrounds is
known {\it a priori\/} to be absent the optimum frequency would clearly
be different.
However as with anisotropy measurements, with multifrequency coverage,
polarized foregrounds can be removed.  

It is also interesting to consider whether the spatial as well as frequency
signature of the polarization can be used to separate foregrounds.
Using angular power spectra for the spatial properties of the foregrounds
is a simple generalization of methods already used in anisotropy work.
For instance, in the case of synchotron emission, if the spatial correlation 
in the polarization follows that of the
temperature itself, the relative contamination will decrease
on smaller angular scales due to its diffuse nature.
Furthermore the peak of the cosmic signal in polarization occurs at 
even smaller angular scales than for the anisotropy.

One could attempt to exploit the additional properties of polarization,
such as its $E$- and $B$-mode nature.
We mentioned above that in a wide class of models where scalars
dominate on small angular scales, the polarization is predicted to be
dominantly $E$-mode (\cite{KamKosSte} 1997 \cite{ZalSel} 1997).  
\cite{Sel} (1997) suggests that 
one could use this to help eliminate foreground
contamination by ``vetoing'' on areas of $B$-mode signal.
However in general one does not expect that the foregrounds will have equal
$E$- and $B$-mode contribution, so while this extra information is valuable,
its use as a foreground monitor can be compromised in certain circumstances. 
Specifically, if a correlation exists between the direction of polarization
and the rate of change (curvature) of its amplitude, the foreground will
populate the two modes unequally. 
Two simple examples: either radial or tangential polarization around a source
with the amplitude of the polarization dropping off with radius, or
polarization parallel or perpendicular to a ``jet'' whose amplitude drops
along the jet axis.
Both examples would give predominantly $E$-mode polarization.

\subsection{Data Analysis} \label{sec:DA}

Several authors have addressed the question of the optimal estimators of
the polarization power spectra from high sensitivity, all-sky maps of the
polarization.  They suggest that one calculate the coefficients of the
expansion in spin-2 spherical harmonics and then form quadratic
estimators of the power spectrum, as the average of the squares 
of the coefficients over $m$, corrected for noise bias as in
the example of Fig.~\ref{fig:sCDM}.

We shall return to consider this below, however before such maps are obtained
we would like to know how to analyze ground based polarization data.
These data are likely to consist of $Q$ and $U$ measurements from tens or
perhaps hundreds of pointings, convolved with an approximately gaussian
beam on some angular scale.  How can we use this data to provide constraints
or measurements of the electric and magnetic power spectra, presumably
averaged across bands in $\ell$?

For a small number of points, the simplest and most powerful way to obtain
the power spectrum is to perform a likelihood analysis of the data.
The likelihood function encodes all of the information in the measurement
and can be modified to correctly account for non-uniform noise, sky coverage,
foreground subtraction and correlations between measurements.
Operationally, one computes the probability of obtaining the measured points
$Q_i \equiv Q(\hhat{n}_i)$ and $U_i \equiv U(\hhat{n}_i)$ assuming a given
``theory'' (including a model for foregrounds and detector noise) and
maximizes the likelihood over the theories.
For our purposes, the theories could be given simply by the polarization
bandpowers in $E$ and $B$ for example, or could be a more ``realistic''
model such as CDM with a given reionization history.
The confidence levels on the parameters are obtained as moments of the
likelihood function in the usual way.
Such an approach also allows one to generalize the analysis to include
temperature information (for the cross correlation) if it becomes
available.

Assuming that the fluctuations are gaussian, the likelihood function is
given in terms of the data and the correlation function of $Q$ and $U$
for any pair of the $n$ data points.
The calculation of this correlation function is straightforward, and
\cite{KamKosSte} (1997) discuss the problem extensively.
Let us assume that we are fitting only one component or have only one
frequency channel.  The generalization to multiple frequencies with a
model for the foreground is also straightforward.  We shall also assume for
notational simplicity that we are fitting {\it only\/} to polarization data,
though again the generalization to include temperature data is straightforward.
The construction is as follows.  We define a data vector which contains
the $Q$ and $U$ information referenced to a particular coordinate system
(in principle this coordinate system could change between different subsets
of the data).
Call this data vector 
\begin{equation}
D = (Q_1,U_1,\ldots,Q_n,U_n) ,
\end{equation}
which has $N=2n$ components.
We can construct the likelihood of obtaining the data given 
a theory once we know the correlation matrix $C_{IJ}$:
\begin{equation}
{\cal L}(D|T) \propto {1\over\sqrt{\det{C}}}
  \exp\left[-{1\over 2}D_I C^{-1}_{IJ} D_J \right]
\label{eqn:likedef}
\end{equation}
All of the theory information is encoded in $C_{IJ}=C_{IJ}^{\rm th}+N_{IJ}$,
where $N_{IJ}$ is the noise correlation matrix, to be provided by the
experiment, and $C_{IJ}^{\rm th}$ is a function of the theory parameters.

All that remains is to compute each element of $C_{IJ}^{\rm th}$ for a
given theory.  Consider a pair of points $i$ and $j$ corresponding to 4
entries of our data vector $D_I$.
Following \cite{KamKosSte} (1997), define $Q'$ and $U'$ as the components of
the polarization in a new coordinate system, where the great arc connecting
$\hhat{n}_1$ and $\hhat{n}_2$ runs along the equator.
Expressions for $\langle Q' Q'\rangle$ and $\langle U' U'\rangle$ in
terms of the $E$ and $B$ angular power follow directly from the definitions
of these spectra and the spin-weighted spherical harmonics.  They can be
found in \cite{KamKosSte} (1997) [their Eqs. (5.9,5.10)].
They also give the flat sky limit of these equations.
Knowing the angle $\phi_{ij}$ about $\hhat{n}_i$ through which we must rotate
our primed coordinate system to return to the system in which our data is
defined we can write
\begin{eqnarray}
Q_i &=&
  Q_i' \cos2\phi_{ij} + U_i'\sin2\phi_{ij} \nonumber\\
U_i &=&
  Q_i'\sin2\phi_{ij} - U_i'\cos2\phi_{ij}
\end{eqnarray}
and similarly for the $j$th element. 
Thus using the known expressions for
$\langle Q' Q'\rangle$ and $\langle U' U'\rangle$ we can calculate
$\langle Q_i Q_j\rangle$, $\langle U_i U_j\rangle$, $\langle Q_i U_j\rangle$
and $\langle U_i Q_j\rangle$ for the $i$th and $j$th pixel and thus all 
of the elements of $C_{IJ}^{\rm th}$.
Substitution of $C_{IJ}^{\rm th}$ into Eq.~(\ref{eqn:likedef}) allows one to
obtain limits on any theory given the data.

Finally let us note that the method outlined here is completely general
and thus can be applied also the the high-sensitivity, all-sky maps which
would result from satellite experiments.  For these experiments, the large
volume of data is however an issue in the analysis pipeline design, as
has been addressed by several authors.  The generalization of the above
analysis procedure to include filtering and compression is straightforward,
and directly analogous to the case of anisotropy, so we will not discuss
it explicitly here.

\section{Future Prospects}
\label{sec:future}

Following a hiatus of some years, experimental efforts to detect CMB
polarization are now underway.  
At the large angular scale, an experiment based in Wisconsin
(\cite{Keaetal}~1997) plans to make a $30$GHz measurement of polarization
on $7^\circ$ angular scales with a sensitivity of a few $\mu $K per pixel.  
The main goal of the experiment would be to look for the signature of
reionization (see Fig.~\ref{fig:sCDM}).  Several groups are currently
considering measuring the signature at smaller scales from recombination.

The {\sl MAP\/} and {\sl Planck\/} missions also plan to measure polarization.
Because of their all sky nature these missions will be the first to be able
to measure the polarization power spectrum and temperature cross correlation
with reasonable spectral sensitivity.
We show an example of the sensitivity to the which {\sl MAP\/} should nominally
obtain in the absence of foregrounds and systematic effects in
Fig.~\ref{fig:sCDM}.
As can be seen, {\sl MAP\/} should certainly detect the $E$-mode of
polarization at medium angular scales, and obtain a significant detection of
the temperature polarization cross-correlation. 
These signatures can be useful for separating adiabatic and isocurvature
models of structure formation as we have seen in \S\ref{sec:reconstruction}.

However the detailed study of the polarization power spectrum will require
the improved sensitivity and expanded frequency coverage of {\sl Planck\/}
for detailed features and foreground removal respectively.
{\sl Planck\/} will have the sensitivity to make a measurement of the
large-angle polarization predicted in CDM models, regardless of the epoch
of reionization if foreground contamination can be removed.  
Likewise, it can potentially measure small levels of $B$-mode polarization. 
The separation of the $E$ and $B$ modes is of course crucial for the isolation
of scalar, vector and tensor modes and so the reconstruction problem in
general.

Clearly, the polarization spectrum represents another gold mine of information in
the CMB.  Though significant challenges will have to be overcome, the prospects
for its detection are bright.

\noindent{\it Acknowledgments:} We thank D. Eisenstein, S. Staggs,
M. Tegmark, P. Timbie and M. Zaldarriaga for useful discussions.
W.H. acknowledges support from the W.M. Keck foundation.

\vskip 1truecm
{\tt \noindent
whu@ias.edu

\noindent
http://www.sns.ias.edu/$\sim$whu/polar/polar.html}
\end{document}